\documentclass{emulateapj}
\usepackage{amssymb, amsmath, epsfig}
\usepackage{fancyvrb}

\newcommand{\bfC}{{\boldsymbol{C}}}

\newcommand{\Trans}{{\cal T}}
\newcommand{\HeIallowedn}{{$1{\rm s}^2~^1 {\rm S}- 1{\rm s} n{\rm p}~^1 {\rm P}^{\rm o}$}}
\newcommand{\HeIallowed}{{$1{\rm s}^2~^1 {\rm S}- 1{\rm s} 2{\rm p}~^1 {\rm P}^{\rm o}$}}
\newcommand{\HeIallowedbeta}{{$1{\rm s}^2~^1 {\rm S}- 1{\rm s} 3{\rm p}~^1 {\rm P}^{\rm o}$}}

\newcommand{\HI}{H{\sc ~i}}

\newcommand{\HeI}{He{\sc ~i}}
\newcommand{\HeII}{He{\sc ~ii}}
\newcommand{\HeIII}{He{\sc ~iii}}

\def\apj{ApJ}
\def\apjl{ApJL}
\def\apjs{ApJS}
\def\aj{AJ}

\def\mnras{MNRAS}

\def\aap{{A\&A}}
\def\prd{PRD}
\VerbatimFootnotes

\begin{document}

%for MNRAS
%\title{The \HeI\ $584~$\AA\ Forest as a Diagnostic of Helium Reionization}
%\author[M. McQuinn \& E. Switzer]{Matthew McQuinn$^1$\thanks{mmcquinn@berkeley.edu} and Eric R. Switzer$^{2}$\\
%$^{1}$ Einstein Fellow; Department of Astronomy, University of California, Berkeley, CA, 94720, USA; mmcquinn@berkeley.edu\\
%$^2$  Kavli Institute for Cosmological Physics, The University of Chicago, Chicago, IL, 60637, USA\\
%}
%\pubyear{2009} \volume{000} \pagerange{1}
%\maketitle\label{firstpage}

%for apj format
\title{The \HeI\ $584~$\AA\ Forest as a Diagnostic of Helium Reionization}
\author{Matthew McQuinn\altaffilmark{1} and Eric R. Switzer\altaffilmark{2}}

\altaffiltext{1} {Einstein Fellow; Department of Astronomy, University of California, Berkeley, CA, 94720, USA; mmcquinn@berkeley.edu}
\altaffiltext{2} {Kavli Institute for Cosmological Physics, The University of Chicago, Chicago, IL, 60637, USA}

\begin{abstract}
We discuss the potential of using the \HeI\ $584~$\AA\ forest to detect and study \HeII\ reionization.  Significant $584~$\AA\ absorption is expected from intergalactic \HeII\ regions, whereas there should be no detectable absorption from low density gas in \HeIII\ regions.   Unlike \HeII\ Ly$\alpha$ absorption (the subject of much recent study), the difficulty with using this transition to study \HeII\ reionization is not saturation but rather that the absorption is weak.  The Gunn-Peterson optical depth for this transition is $\tau \sim 0.1 \,x_{\rm HeII} \,\Delta^2 \, [(1+z)/5]^{9/2}$, where $x_{\rm HeII}$ is the fraction of helium in \HeII\ and $\Delta$ is the density in units of the cosmic mean.   In addition, \HeI\ $584~$\AA\ absorption is contaminated by lower redshift \HI\ Ly$\alpha$ absorption with a comparable flux decrement.  We estimate the requirements for a definitive detection of redshifted \HeI\ absorption from low density gas ($\Delta \approx 1$), which would indicate that \HeII\ reionization was occurring.  We find that this objective can be accomplished (using coeval \HI\ Ly$\alpha$ absorption to mask dense regions and in cross correlation) with a spectral resolution of $10^4$ and a signal-to-noise ratio per resolution element of $\sim 10$.  Such specifications may be achievable on a few known $z\sim 3.5$ quasar sightlines with the Cosmic Origins Spectrograph on the Hubble Space Telescope.  We also discuss how \HeI\ absorption can be used to measure the hardness of the ionizing background above $13.6$~eV.
\end{abstract}

\keywords{cosmology: theory -- cosmology: intergalactic medium -- quasars: absorption lines}

\section{Introduction}

Starlight produced by the first galaxies is the leading candidate for ionizing the hydrogen as well as singly ionizing the helium at $z\sim 6$.  It takes a harder source of radiation to doubly ionize the helium, so the reionization of this species is likely deferred until $z\sim 3$ when quasars produce a sufficient hard UV background \citep{madau99, furlanetto08, mcquinn09}.  However, the helium could have been doubly ionized at nearly the same cosmic time that hydrogen was reionized if more exotic sources ionized the hydrogen, such as the first generation of metal-free stars \citep{bromm01, venkatesan03, tumlinson04} or miniquasars  \citep{madau99, volonteri09}.  In a third potential scenario, early sources doubly ionized the helium and then shut off.  Afterward, the \HeII\ recombined such that quasars could again reionize it at $z\sim 3$ \citep{wyithe03, venkatesan03}.

% It takes a hard source of radiation to doubly ionize the helium.  In this picture, the radiation from quasars doubly ionized the intergalactic helium at $z\sim 3$ \citep{madau99, furlanetto08, mcquinn09}.

If \HeII\ reionization were completing at $z \sim 3$, an epoch for which there are numerous observations of the intergalactic medium (IGM), it should be an easier task to definitively detect this process compared to detecting $z \gtrsim 6$ reionization processes.  Furthermore, if \HeII\ reionization were ending at $z \sim 3$, it should have significantly affected the temperature of the intergalactic gas and the ultraviolet radiation background.  These motivations, along with recent additions to the Hubble Space Telescope (HST), have inspired a significant effort of late to understand the signatures and the detection prospects of \HeII\ reionization \citep{furlanetto08, mcquinn09, lidz09, bolton09, mcquinn09b, dixon09, furlanetto09, syphers09, syphers09b, mcquinn3He}.

Three separate observations of the $z\sim 3$ IGM suggest that \HeII\ reionization was ending around this redshift:  First, several studies have measured the temperature of the intergalactic gas from the widths of the narrowest lines in the \HI\ Ly$\alpha$ forest, and the majority of these studies have found evidence for an increase in the IGM temperature of $\sim 10^4$~K between $z \approx 4$ and $z \approx 3$, before a decline to lower redshift \citep{schaye00, ricotti00, lidz09}.  These trends have been attributed to the heating from \HeII\ reionization.   Second, observations of \HeII\ Ly$\alpha$ absorption from gas at $2.8 < z < 3.3$ show tens of comoving Mpc (cMpc) regions with no detected transmission \citep{reimers97, heap00}, which may indicate that \HeII\ reionization was not complete.  Thirdly, \citet{songaila98} and \citet{agafonova07} detected evolution in the column density ratios of certain highly ionized metals at $z \approx 3$, which they argued was due to a hardening in the ionizing background around $50~$eV and, thus, the end of \HeII\ reionization.  

However, the interpretations of all of these indications for \HeII\ reionization are controversial.  Temperature measurements of the IGM are difficult, and not all measurements detected the aforementioned trends.  It is often argued that \HeII\ Ly$\alpha$ absorption saturates at \HeII\ fractions that are too small ($\sim 10^{-3}$ at the mean density) to study \HeII\ reionization (although, see \citealt{mcquinn09b}).  Lastly, inferences from metal lines require significant modeling, and studies have reached different conclusions regarding the degree of their evolution at $z\sim 3$ \citep{boksenberg03}.

This paper discusses intergalactic absorption by the \HeIallowedn\ transitions of \HeI\ as an unsaturated observable of \HeII\ reionization.   We primarily focus on the strongest and longest wavelength of these absorption lines, the \HeI\ \HeIallowed, $584~$\AA\ line.  For a given optical depth in the \HI\ Ly$\alpha$ forest,  the amount of absorption in the \HeI\ forest can be directly estimated if both the fraction of helium that is \HeII\ and the ratio of the \HI\ and \HeI\ photoionization rates are known.  In addition, because the \HeI\ ionization edge is relatively close to that of hydrogen, the photoionization rate that \HeI\ experiences is similar to this for hydrogen.  Thus, the most important determinant of the amount of \HeI\ $584$~\AA\ absorption is the \HeII\ fraction.

Other studies have discussed the \HeI\ forest, but did not focus on its usefulness as a probe of helium reionization.  \citet{tripp90} was the first to discuss intergalactic absorption by the \HeI\ $584~$\AA, and this study attempted unsuccessfully to detect this absorption in the spectrum of a $z=1.7$ quasar.  \citet{reimers93} targeted \HeI\ absorption from $z\approx 2$ \HI\ Lyman-limit systems, and reported the first (and at present only) detection of intergalactic \HeI\ $584~$\AA\ absorption.  Finally, \citet{miralda92} showed that the \HeI\ forest could be a useful probe of the hardness of the ultraviolet background, and they focused on using this absorption to rule out a particular nonstandard model for the dark matter.  \citet{santos03} followed up on this idea, arguing that the $z\sim 5$ \HeI\ forest could be a useful diagnostic of the hardness of the ultraviolet background after hydrogen reionization.

As with the \HeII\ Ly$\alpha$ transition at $304~$\AA, the \HeI\ $584~$\AA\ transition falls blueward of the hydrogen Lyman limit and, therefore, is subject to continuum absorption by hydrogen, in this case from \HI\ systems with $z > z' \equiv 3.2 \, (1 +z_{\rm HeI})/5 -1$.  This continuum absorption may even be worse in terms of obscuring the \HeI\ $584~$\AA\ forest compared to the \HeII\ Ly$\alpha$ forest because this spectral region is more affected by higher redshift \HI\ continuum absorbers.  It is unlikely that there are more than a handful of quasar sightlines at $z>4$ with sufficient near ultraviolet (NUV) flux for detection with the present generation of instruments (and almost certainly not at $z>6$, during \HeI\ reionization; \citealt{santos03}).  Therefore, our focus is on applications of \HeI\ absorption at $z\lesssim 4$.

We have performed a cursory search for candidate \HeI\ sightlines among the relatively few published HeII Lya forest spectra that extend redward to $584 \, (1+z_{\rm QSO})~$\AA.  Of note,  QSO~OQ~172 ($z=3.54$) has $F_\lambda \approx 2 \times 10^{-16}~$erg~s$^{-1}$~cm$^{-2}$~\AA$^{-1}$ at $584 \, (1+z) ~ $\AA\ \citep{lyons95}, and QSO~0055-269 ($z=3.67$) has $F_\lambda \approx 1 \times 10^{-16}$ (Gabor Worseck, priv. com.).  In fact, we were surprised to find that most \HeII\ sightlines in our search had significant flux in the relevant band.  

Many of the existing \HeII\ sightlines had been selected by their far ultraviolet flux.  NUV selection would be more optimal for identifying candidate \HeI\ forest sightlines.  For example, HS~1140~+3508 ($z=3.15$) is obscured in the far ultraviolet by a Lyman-limit system, but has a NUV flux of $F_\lambda \approx 2 \times 10^{-16}$ (Gabor Worseck, priv. com.). 

Another difficulty with the \HeI\ forest is that line absorption from foreground systems can contaminate the \HeI\ forest, the most important of which is \HI\ Ly$\alpha$ absorption.  \HI\ Ly$\alpha$ absorption from a system with a redshift of $z_{\rm HI, 1216} = 2.4 \, (1 +z_{\rm HeI, 584})/5 -1$ falls directly on top of the \HeI\ absorption from a redshift of $z_{\rm HeI, 584}$.  Fortunately, the Ly$\alpha$ forest is quite thin at relevant $z_{\rm HI, 1216}$ (with a flux decrement of several percent), and we find that both this contaminant's mean transmission and variance tend to be comparable to that of the \HeI\ forest.  Also, the \HeI\ forest correlates strongly with the coeval \HI\ Ly$\alpha$ forest, which allows it to be extracted in cross correlation despite this contamination.

This study is timely because the HST reservicing mission installed the Cosmic Origins Spectrograph (COS) in May 2009.  COS is capable of measurements of HeI $584~$\AA\ absorption at redshifts relevant to \HeII\ reionization ($2.8 \lesssim z < 4.5$), and ground-based spectrographs can cover higher redshifts ($z \gtrsim 4.3$, corresponding to $\gtrsim 3100$~\AA).  COS is able to achieve higher signal-to-noise ratios than previous instruments in the ultraviolet.  A $60~$hr exposure with COS would achieve a signal-to-noise ratio of $10$ at ${\it R} \approx 30,000$ for a flux of $F_\lambda =  2 \times 10^{-16}~$erg~s$^{-1}$~cm$^{-2}$~\AA$^{-1}$ (the flux of QSO~OQ~172 and HS~1140~+3508) at $2600~$\AA\ (\HeI\ absorption from $z=3.5$).  A $40~$hr exposure with COS would achieve a signal-to-noise ratio of $10$ for this flux at ${\it R} \approx 20,000$ at $2300~$\AA\ ($z=2.9$).\footnote{http://etc.stsci.edu/webetc/.  As of mid-January 2010, the dark current for the COS NUV gratings is $5$ times higher than specifications (and than what is assumed by this ETC), and, thus, a significantly longer observation is required unless this issue is resolved.}  

This paper is organized as follows.  Section~\ref{sec:forest} discusses the physics of the \HeI\ forest.   Section~\ref{sec:calculations} contrasts simulated spectra for this forest under different assumptions regarding the ionization state of the helium.  Section~\ref{sec:tests} quantifies the spectral quality an observation must achieve to verify whether \HeII\ reionization was occurring from the \HeI\ forest.  Appendix~\ref{sec:hardness} outlines how to measure the hardness of the ionizing background with \HeI\ absorption, and Appendix~\ref{ap:SNRcc} derives formulae for the significance with which \HeI\ absorption can be detected in cross correlation with the coeval Ly$\alpha$ forest.  

This paper assumes a flat $\Lambda$CDM cosmology with $h =0.7$, $\Omega_b = 0.046$, $\Omega_m = 0.28$, $\sigma_8 = 0.82$, $n_s = 1$, and $Y_{\rm He} = 0.24$, consistent with recent measurements \citep{komatsu08}.  However, the simulation used to calculate the Ly$\alpha$ forest spectra at $z < 1.5$, the D5 simulation in \citet{springel03}, assumes a slightly different cosmology, with the most notable differences being $\Omega_b = 0.04$ and $\sigma_8 = 0.9$.  The photoionization and recombination rates used in this study are from \citet{hui97}.  

\section{The \HeI\ Forest}
\label{sec:forest}

In photoionization equilibrium,  the \HeI\ fraction is determined by the relation
\begin{equation}
x_{\rm HeI} \approx  \frac{\alpha_{\rm HeI} \, n_e}{\Gamma_{\rm HeI}} \, x_{\rm HeII}.
\label{eqn:xHeI}
\end{equation}
Photoionization equilibrium is a good approximation because the timescale to reach equilibrium ($\Gamma_{\rm HeI}^{-1} \sim \Gamma_{\rm HI}^{-1} \approx 3 \times 10^4$~yr at $z\sim 3$; e.g., \citealt{faucher08}) was much shorter than all other relevant timescales.
  Here, $\alpha_{X}$,  $n_{X}$, $x_X$, and $\Gamma_{X}  \equiv \int_{E^{\rm ion}_X}^\infty (dE/E) \, \sigma_X(E) \, J(E)$ are respectively the Case A recombination coefficient,\footnote{Case A is most relevant for \HI\ and \HeI\ because this paper's focus is on intergalactic systems in which these species are highly ionized.} the number density,  the ionization fraction, and the photoionization rate for ion $X$ (or subscript $e$ for electrons), and $E^{\rm ion}_X$ and $\sigma_X$ are the ionization potential and the photoionization cross section.  Lastly, $J(E)$ is the incident specific intensity integrated over solid angle.  We will sometimes imprecisely write $x_{\rm HeII} = 1$ for a gas parcel even though a small fraction ($\sim 10^{-5}\, \Delta_b$) of the helium is in $x_{\rm HeI}$.

If the density and $\Gamma_{\rm HeI}$ are known (since $n_e$ is effectively known in the IGM after hydrogen reionization up to density), a measurement of $x_{\rm HeI}$ can be used to determine $x_{\rm HeII}$ (equation~\ref{eqn:xHeI}).  Fortunately, coeval \HI\ Lyman-series absorption provides an estimate for the density of a \HeI\ absorber.  In addition, the \HeI\ photoionization rate (like this for the \HI) is expected to have been essentially spatially independent owing to the long mean free path of \HeI-ionizing photons ($l_{\rm HeI}$)  and the large number of sources in a volume of $\sim l_{\rm HeI}^3$.  It is estimated that $l_{\rm HeI} > l_{\rm HI} \sim 300~$cMpc at $z \sim 3$ \citep{madau99, faucher08, prochaska09}.  Therefore, $\Gamma_{\rm HeI}$ was effectively just a single number in all of the IGM. 
% Section~\ref{sec:forestGamma} discusses the uncertainty in theoretical estimates of $\Gamma_{\rm HeI}$.  

This paper focuses primarily on the longest wavelength allowed ground-state transition for \HeI, the \HeIallowed, $584~$\AA\ transition.  Its oscillator strength is $3.8$ times larger than the next strongest transition of \HeI, the \HeIallowedbeta, $537~$\AA\ transition.  \HeI\ $584~$\AA\ absorption is observable in a suitable quasar absorption spectrum between the wavelengths of $584 \, (1+z_{\rm QSO})~$\AA\ and $304\, (1+z_{\rm QSO})~$\AA, where the \HeII\ Ly$\alpha$ forest begins.  This wavelength range corresponds to \HeI\ absorption from $0.5 \, z_{\rm QSO} <z < z_{\rm QSO}$.\footnote{\HeI\ continuum absorption by intervening systems will produce breaks in the spectrum starting blueward of $504 \, (1 + z_{\rm QSO})~$\AA\ (the redshifted \HeI\ limit), but these discontinuities should be minimal ($\lesssim 10\%$) for sightlines selected against having strong \HI\ continuum absorption.}

To redshift across the $584~$\AA\ transition, a photon experienced the optical depth from gas in the Hubble flow of
\begin{eqnarray}
\tau_{\rm HeI, 584}^{GP} &\approx& 0.07 \, x_{\rm HeII} \; \Delta_b^{2} \; T_4^{-0.7} \left(\frac{10^{-12} \, {\rm s}^{-1}}{\Gamma_{\rm HeI}} \right) \left(\frac{1+z}{5} \right)^{{9}/{2}},  \label{eqn:584GP} \nonumber \\
&\approx& 0.025 \, x_{\rm HeII} \,  \frac{\Gamma_{\rm HI}}{\Gamma_{\rm HeI}}\; \tau_{\rm HI, 1216}^{\rm GP} ,
\label{eqn:584GP2}
\end{eqnarray}
(what is termed the \HeI\ Gunn-Peterson optical depth; \citealt{gunn65,tripp90}) where
\begin{equation}
\tau_{\rm HI, 1216}^{\rm GP} \approx 2.7 \, \Delta_b^{2} \; T_4^{-0.7}  \left(\frac{10^{-12}\, {\rm s}^{-1}}{\Gamma_{\rm HI}} \right) \left(\frac{1 +z}{5}\right)^{{9}/{2}}
\label{eqn:tauHIGP}
\end{equation} 
is the \HI\ Ly$\alpha$ Gunn-Peterson optical depth, $\Delta_b$ is the gas density in units of the cosmic mean, $T_4$ is the temperature in units of $10^4~$K,  and the $-0.7$ exponent owes to the temperature dependence of the recombination coefficient (cf. equation~\ref{eqn:xHeI}).  
%The temperature of the $z\sim 3$ IGM at mean density is measured to have been $\approx 2\times 10^4~$K (e.g., \citealt{lidz09}), but regions in which the \HeII\ has not been reionized (where \HeI\ absorption occurs) were likely colder by $\sim 10^4~$K (e.g., \citealt{mcquinn09}).  In addition, denser regions should have had higher temperatures, with the temperature scaling as $\Delta_b^\beta$ where $0 < \beta < 0.6$ \citep{hui97}.   

Overdensities of a few and greater at $z \sim 3$ had decoupled from the Hubble flow and were collapsing or had collapsed.  In these regions, the Gunn-Peterson optical depth no longer describes the absorption.  Instead, such regions appear as distinct absorption lines with widths of $10$s of km~s$^{-1}$, and their optical depth in \HeI\ $584~$\AA\ is
\begin{eqnarray}
\tau_{\rm HeI, 584}^{N_{\rm HI}} &\approx& 0.11 \, x_{\rm HeII} \frac{\Gamma_{\rm HI}}{\Gamma_{\rm HeI}} \left(\frac{30 \,{\rm km \, s^{-1}}}{\Delta v_{\rm HeI}} \right) \left(\frac{N_{\rm HI}}{10^{14} \, {\rm cm}^{-2}} \right), \nonumber \\
&\approx& 0.025 \, x_{\rm HeII} \,  \frac{\Gamma_{\rm HI}}{\Gamma_{\rm HeI}} \left( \frac{{\Delta v}_{\rm HI}}{{\Delta v}_{\rm HeI}} \right) \, \tau_{\rm HI, 1216}^{N_{\rm HI}},
\label{eqn:tauNHI}
\end{eqnarray}
where $\tau_{\rm HI, 1216}^{N_{\rm HI}}$ is the \HI\ Ly$\alpha$ optical depth, and we have approximated the line profile as a tophat with velocity width $\Delta v_X$ (and $1 \leq {{\Delta v}_{\rm HI}}/{{\Delta v}_{\rm HeI}} \leq 2$, with $2$ being the limit of pure thermal broadening).  Systems with $N_{\rm HI} > 10^{14} \, {\rm cm}^{-2}$ are common in the Ly$\alpha$ forest.  Each sightline intersects $\sim 100$ such systems between $z=3$ and $z=4$ (e.g., \citealt{press93}).

\subsection{\HeI\ Photoionization Rate}
\label{sec:forestGamma}

The \HeI\ $584$~\AA\ optical depth depends on the value of $\Gamma_{\rm HeI}$ in addition to the field of interest, $x_{\rm HeII}$.  The calculations in this paper assume $\Gamma_{\rm HeI} = \Gamma_{\rm HI}$, and $\Gamma_{\rm HI}$ is chosen to match measurements of the \HI\ Ly$\alpha$ mean transmission (requiring $\Gamma_{\rm HI} \sim 10^{-12}~$s$^{-1}$; e.g., \citealt{faucher08}).   This choice is motivated by ultraviolet background models, which find $\Gamma_{\rm HeI} \approx \Gamma_{\rm HI}$ at $z = 2-4.5$ \citep{haardt96, faucher09}.  The calculation of $\Gamma_{\rm HI}$ and $\Gamma_{\rm HeI}$ in ultraviolet background models requires two inputs: (1) a model for the source emissivity at frequencies near the ionization potential for \HI\ and \HeI, and (2) the measured \HI\ column density distribution, which determines the attenuation by \HI\ continuum absorption.  %(The other possible ingredient, \HeI\ continuum absorption, is unimportant as explained below.)

The \citet{faucher09} model predicts $\Gamma_{\rm HeI} = 0.8-1.0~\Gamma_{\rm HI}$ at $2 < z < 5$.  This model assumes that both stars and quasars contributed to the metagalactic ultraviolet output, with an increasingly important stellar contribution with redshift at $z \gtrsim 2.5$.  This model uses the result of the stellar population synthesis code PEGASE for the spectral index of starlight between the \HeI\ and \HeII\ ionization edges.  This code yields a similar value for stars ($\alpha \approx -1$, defined as the power-law slope of the specific intensity; \citealt{kewley01}) to what is expected for quasars ($\alpha \approx -1.6$; \citealt{telfer02}).  These similar $\alpha$ result in the near constancy of  $\Gamma_{\rm HeI}(z)$ in the \citet{faucher09} model.  Given these input $\alpha$, \HI\ continuum absorption then resulted in the spectral index of the average radiation background being hardened above $13.6~$eV by roughly $3 \, (\beta -1)$ over the input $\alpha$ for $\beta < 2$, where $\beta$ is the power-law index of the \HI\ column-density distribution.  The background model of \citet{faucher09} assumes $\beta \approx 1.5$, resulting in a hardening of $\Delta \alpha \approx 1.5$. 

However, there is modeling uncertainty in the input $\alpha$ for stars that stems from uncertainty in the spectrum of Wolf-Rayet stars \citep{kewley01}, in the average stellar metallicity, and in the stellar initial mass function.  If the spectral index of stars were significantly softer, the ratio $\Gamma_{\rm HeI}/\Gamma_{\rm HI}$ would be smaller than in the fiducial model in \citet{faucher09}.   In the extreme example that the stellar contribution dominated the ionizing background and had a spectral index of $-3$ instead of $-1$, $\Gamma_{\rm HeI}$ would have been reduced relative to $\Gamma_{\rm HI}$ by a factor of $\approx 3$.  This boosts $\tau_{\rm HeI, 584}$ by a factor of $3$.  In addition, there is uncertainty in $\beta$ at the $\pm 0.3$-level \citep{prochaska10} that affects $\Gamma_{\rm HeI}/\Gamma_{\rm HI}$ at the factor of two-level.

Previous calculations of the ionizing background included only \HI\ continuum absorption to calculate (given a source emissivity model) the ionizing background between $E^{\rm ion}_{\rm HI}$ and $0.75 \,E^{\rm ion}_{\rm HeII}$.  These models ignored \HeI\ continuum absorption.  This approximation may have affected their estimate for $\Gamma_{\rm HeI}$ because the \HeI\ continuum absorption above $E^{\rm ion}_{\rm HeI}$ is comparable to that of \HI\ for optically thin systems.  Namely, for gas of primordial composition in photoionization equilibrium 
\begin{equation}
\tau^{\rm cont}_{\rm HeI}(E^{\rm ion}_{\rm HeI})  \approx 0.5 \,x_{\rm HeII}  \,\frac{\Gamma_{\rm HI}}{ \Gamma_{\rm HeI}}\, \tau^{\rm cont}_{\rm HI}(E^{\rm ion}_{\rm HeI}),
\label{eqn:continuum}
\end{equation}
where $\tau^{\rm cont}_X(E)$ is the continuum absorption of ion $X$ at energy $E$.   

However, most of the absorption of background photons with energy $\approx E^{\rm ion}_{\rm HeI}$ at $z\sim 3$ derived from systems with $N_{\rm HI} \approx 10^{18}~$cm$^{-2}$ (for which $\tau^{\rm cont}_{\rm HI}(E^{\rm ion}_{\rm HeI}) \approx 1$).  These systems were much more self-shielded to \HI\ Lyman-limit photons than to those at the the \HeI\ ionization edge.  Thus, the effective value of $\Gamma_{\rm HI}/ \Gamma_{\rm HeI}$ in equation~(\ref{eqn:continuum}) was much lower in these dense absorbers, resulting in \HI\ continuum absorption dominating over that of \HeI\ at $\approx E^{\rm ion}_{\rm HeI}$.  We estimate in calculations not included that the break at $E^{\rm ion}_{\rm HeI}$ owing to \HeI\ continuum absorption in the spatially-averaged UV background at $z\sim 3$ amounted to only a couple percent.  This justifies our use of a background model that does not include \HeI\ continuum absorption.

\section{Calculations }
\label{sec:calculations}

\begin{figure*}
\begin{center}
{\epsfig{file=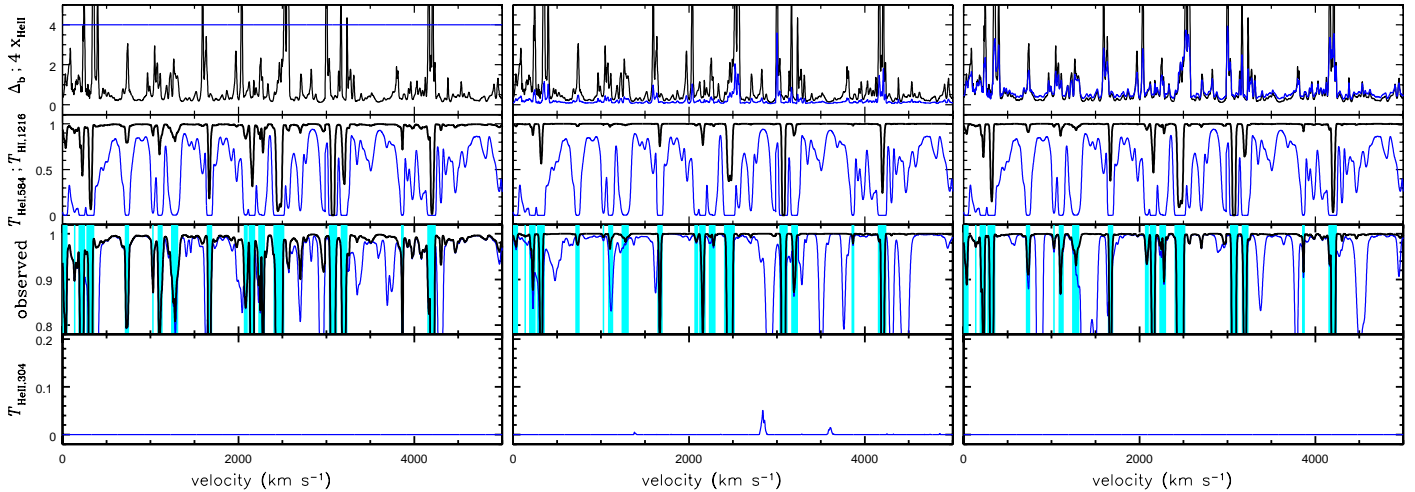, width=19cm}}
\end{center}
\caption{Plot of different quantities from a simulated skewer through the IGM at $z=4$ with length $\approx 60$~cMpc ($\Delta z =  0.08$).  The three large panels make different assumptions about the ionization state of the \HeII.  The leftmost panel assumes $\bar{x}_{\rm HeII} = 1$, the middle that the \HeII\ is kept ionized by the background $\Gamma_{\rm HeII} = 5\times 10^{-16}$~s$^{-1}$, and the right the same but $\Gamma_{\rm HeII} = 10^{-16}$~s$^{-1}$.  For the latter two cases, self-shielding by dense regions is included with an approximate radiative transfer method that is described in the text.  In each larger panel, the top subpanel shows $\Delta_b$ (black curve) and $4 \times x_{\rm HeII}$ (blue curve), the second subpanel down shows the transmission in the \HeI\ $584 \, $\AA\ (black curve) and \HI\ $1216\,$\AA\ (blue curve) forests, and the third subpanel down zooms in on the \HeI\ $584 \, $\AA\ transmission (black curve).  The third panel also shows the observed transmission (blue curve), which, in addition to \HeI\ $584 \, $\AA\ absorption, includes the contamination from foreground \HI\ Ly$\alpha$ absorption.  The highlighted regions represent denser gas in which $\tau_{\rm HI, 1216} > 3$ in the coeval Ly$\alpha$ forest.  Lastly, the bottom subpanel zooms in on the transmission in the \HeII\ Ly$\alpha$ forest.  \label{fig:HeIforest_z4}}
\end{figure*}

\begin{figure*}
\begin{center}
{\epsfig{file=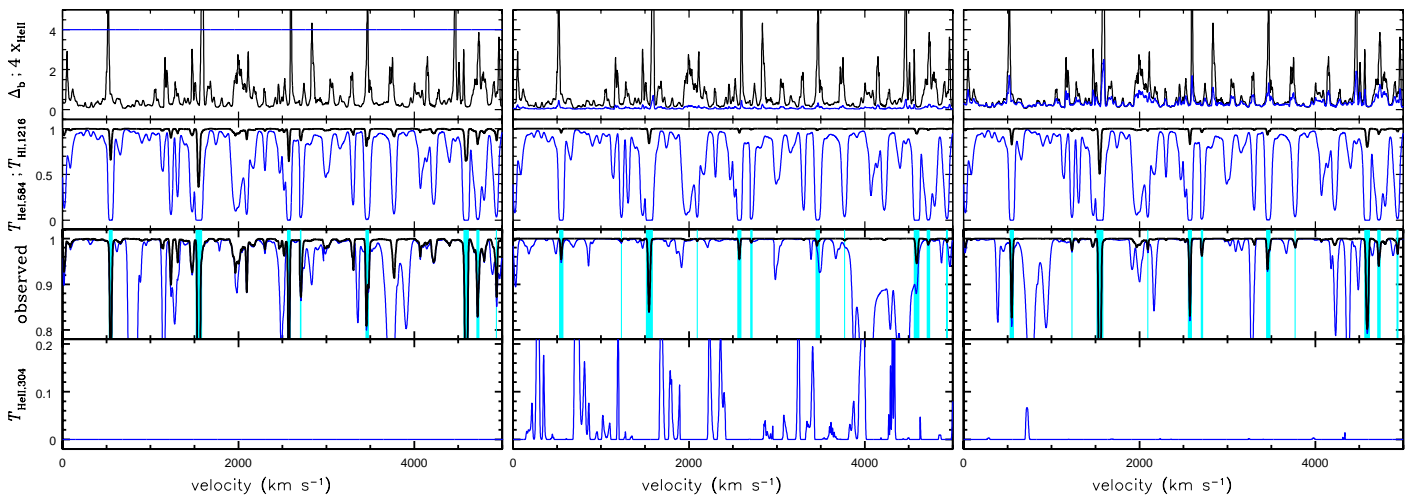, width=19cm}}
\end{center}
\caption{Same as Figure~\ref{fig:HeIforest_z4}, but at $z=3$. \label{fig:HeIforest_z3}}
\end{figure*}

Figures~\ref{fig:HeIforest_z4}~and~\ref{fig:HeIforest_z3} show calculations of the \HeI\ forest at $z =4$ and $z=3$, respectively.  These calculations use a randomly selected skewer of length $\approx 60~$cMpc ($\Delta z \approx 0.1$) through the $25~$cMpc/$h$, $2\times512^3$ particle SPH simulation described in \citet{lidz09}.  The D5 simulation in \citet{springel03} is also used for calculating the low-z Ly$\alpha$ contamination.\footnote{The two simulations were used because the \citet{lidz09} simulation was not run to low enough redshifts to use to calculate the low-redshift Ly$\alpha$ forest contamination.  Both simulations were run to $z=2$.  We have compared calculations of the $z\sim 3$ absorption in these two simulations (which have a mass resolution that differs by a factor of $10$) to verify convergence.}   The mean transmission in the \HI\ Ly$\alpha$ forest has been rescaled in these calculations by adjusting $\Gamma_{\rm HI}$ to match the observed values.  In particular, we use the $\Gamma_{\rm HI}$-values of \citet{faucher08b} for $z >2$ and those in \citet{kirkman07} for $z <2$.  The \HeI\ $584~$\AA\ forest is calculated in post-processing with the conservative assumption that $\Gamma_{\rm HeI} = \Gamma_{\rm HI}$ (Section \ref{sec:forestGamma}).

The three large panels in Figures~\ref{fig:HeIforest_z4}~and~\ref{fig:HeIforest_z3} make different assumptions regarding the ionization state of the helium.  The leftmost assumes that the \HeII\ fraction is given by ${x}_{\rm HeII} = 1$ everywhere, the middle that the helium is mostly \HeIII\ and that the fraction in \HeII\ versus \HeIII\ is determined by a weak \HeII-ionizing background with $\Gamma_{\rm HeII} = 5\times10^{-16}~$s$^{-1}$, and the rightmost assumes an even weaker background with $\Gamma_{\rm HeII} = 10^{-16}~$s$^{-1}$. 
The value $\Gamma_{\rm HeII} = 5\times10^{-16}~$s$^{-1}$ is approximately the upper bound that has been set from observations of the most opaque regions in the \HeII\ Ly$\alpha$ forest at $z \approx 3$ \citep{mcquinn09b}.  %These $\Gamma_{\rm HeII}$-values are $10^3-10^4$ times smaller than the background that was keeping the hydrogen ionized at these redshifts.
 	
The densest absorbers may become self-shielded to \HeII-ionizing photons and can experience an even smaller $\Gamma_{\rm HeII}$.  Self-shielding is included in these calculations with a simplified $1$-D algorithm for the radiative transfer.   This algorithm assumes that regions with $\Delta_b < 1$ experience the quoted $\Gamma_{\rm HeII}$.  For the denser regions, this algorithm locates segments along a given $1$-D skewer that are bounded by $\Delta_b = 1$.  Each segment is composed of tens to hundreds of grid cells, each with width $\approx 1~$km~s$^{-1}$.

For each overdense segment, our scheme attenuates the ionizing background (half of which is assumed to enter from each side) based on the amount and distribution of \HeII.   The scheme starts with the assumption that $x_{{\rm HeII}, {\it i}} = 1$ and $J_{i}(E) = J(E)^{\rm void} \,\exp[-\sigma_{\rm HeII}(E) \, N_{{\rm HeII}, {\it i}}]$ for all cells across the segment, where $i$ labels the cell number,  and $N_{{\rm HeII}, {\rm i}}$ and $J_{i}(E)$ are respectively the \HeII\ column density and incident intensity to cell $i$ from $< i$.  Similarly, there is a contribution to $J_{i}(E)$ from $>\,i$.  Next, the scheme iterates to converge to $\Gamma_{{\rm HeII}, {\it i}}$ and $x_{{\rm HeII}, {\it i}}$, assuming photoionization equilibrium.  For our calculations, $J(E)^{\rm void}$ is set to have a spectral index of $-1.5$, as expected for quasars.
%\footnote{This algorithm will result in an overestimate of the amount of self-shielding because (1) it is likely that the spectral index of the incident radiation is hardened by attenuation prior to reaching an absorber, (2)  typically there will be directions for which the \HeII\ column density of an absorber is smaller than along the chosen $1$-D skewer, and (3) due to the 1-D nature of our calculation, a dense absorber with \HeII\ continuum optical depth greater than unity spuriously results in the regions with $\Delta_b \approx 1$ at the edge of the absorber experiencing a $\Gamma_{\rm HeII}$ that is reduced by a factor of $2$ relative to the mean.}

 \HeII\ Ly$\alpha$ forest sightlines show that $\Gamma_{\rm HeII}$ fluctuates wildly on scales of $\gtrsim10~$cMpc \citep{zheng04, fechner07}.  We do not attempt to model these fluctuations here, but will comment on how they could affect our conclusions.  These fluctuations will spatially modulate the number of dense self-shielding regions compared to the uniform-$\Gamma_{\rm HeII}$ case.
  However, because the \HeI\ forest is sensitive to $x_{\rm HeII} \sim 1$, $\Gamma_{\rm HeII}$ fluctuations are less of a concern for studying \HeII\ reionization with the \HeI\ forest compared to with the \HeII\ Ly$\alpha$ forest (which is sensitive to $x_{\rm HeII} \ll 1$).  %Furthermore, previous measurements these fluctuations from the \HeII\ Ly$\alpha$ forest at $z\approx 2.8$ were only sensitive to $\Gamma_{\rm HeII}$-values that are a factor of $\approx 10$ larger than the value used in our weakest background model. 

The top subpanels in Figures~\ref{fig:HeIforest_z4}~and~\ref{fig:HeIforest_z3} show the value of $x_{\rm HeII}$ that results from this algorithm, where Hubble's law is used to relate position to velocity.  The blue curve is $4 \times x_{\rm HeII}$ and the black curve is $\Delta_b$.  The value of $x_{\rm HeII}$ can be quite large for the weak backgrounds that are assumed.   For $\Delta_b = 1$, $T= 10^4~$K, and $\Gamma_{\rm HeII} = 5 \times 10^{-16}~$s$^{-1}$ ($1 \times 10^{-16}~$s$^{-1}$), $ x_{\rm HeII}$ equals $0.06$ ($0.25$) at $z= 3$. This number becomes $0.11$ ($0.39$) at $z=4$.

The bottom subpanels in each larger panel in Figures~\ref{fig:HeIforest_z4}~and~\ref{fig:HeIforest_z3} zoom in on the residual transmission in the \HeII\ Ly$\alpha$ forest (absorption at wavelengths of $304 \, [1+z] \, $\AA).  Note that in all of the cases there is minimal transmission in this forest because the \HeII\ Ly$\alpha$ transition saturates for $x_{\rm HeII} \sim 10^{-3} \, \Delta_b^{-1}$.  There is only a detectable amount of transmission in the case with $\Gamma_{\rm HeII} = 5\times 10^{-16}~$s$^{-1}$ and $z=3$ (middle panel, Fig.~\ref{fig:HeIforest_z3}).  Because \HeII\ reionization was patchy, regions of transmission in the \HeII\ Ly$\alpha$ forest can occur even if reionization was not complete.  In contrast, \HeI\ absorption has the potential to reveal whether those opaque neighboring regions were in fact \HeII\ regions because it is sensitive to $x_{\rm HeII} \sim 1$.
%\footnote{Note that the \HeII\ Ly$\alpha$ forest is sensitive to the temperature of the IGM, which is not accurately captured in these simulations.  Increasing the temperature decreases the amount of \HeII, resulting in increased transmission, but it also increases the Jeans scale, which acts to decrease the transmission.}

The middle subpanels in Figures~\ref{fig:HeIforest_z4}~and~\ref{fig:HeIforest_z3} show the transmission in the \HI\ Ly$\alpha$ forest (blue curves) and \HeI\ $584~$\AA\ forest (black curves).  These panels demonstrate that the transmission in the \HI\ and \HeI\ forests is highly correlated.  The \HeI\ absorbers that correspond to the deepest \HI\ Ly$\alpha$ forest lines are the most visible, and the weaker \HeI\ lines (systems that are not dense enough to self-shield) disappear in the cases in which the helium is mostly doubly ionized.
%\footnote{The temperature of the IGM at $3 < z < 4$ is on the low side in the \citet{lidz09} simulation compared to measurements.  In this simulation, $T \approx 10-15~$kiloK at the mean density rather than the measured value of $T = 20\pm 5~$kiloK (e.g., \citealt{lidz09}).  Prior to the \HeII\ being ionized, it is more likely that $T$ is in this $10-15~$kiloK range \citep{mcquinn09}, but in intergalactic \HeIII\ regions the simulation's values for $T$ are likely too small.  The simulation's low temperature in \HeIII\ regions increases the difficulty of distinguishing between the models in which \HeII\ reionization has happened and those in which it has not (because $x_{\rm HeII} \sim T^{-0.7}$ in photoionization equilibrium).}  

The third subpanel down in each larger panel in Figures~\ref{fig:HeIforest_z4}~and~\ref{fig:HeIforest_z3} zooms in on the the \HeI\ $584 \, $\AA\ transmission (solid black curves).  These subpanels also include mock realizations of the observed spectrum at these wavelengths (i.e., \HeI\ $584 \, $\AA\ plus foreground \HI\ Ly$\alpha$ absorption; solid blue).  Each panel uses a different skewer through the IGM to calculate this foreground absorption.
The highlighted regions in these panels represent the locations with $\tau_{\rm HI, 1216} > 3$ for absorption coeval to that of the \HeI.  Outside of the highlighted regions, the amount of \HeI\ absorption is substantially different between the three cases in each figure.  Trace amounts of absorption remain in essentially just the $x_{\rm HeII} = 1$ case.  A detection of \HeI\ absorption in these regions would indicate that \HeII\ reionization was occurring.  The next section quantifies the prospects for detecting this absorption.

\begin{figure}
\begin{center}
\rotatebox{-90}{\epsfig{file=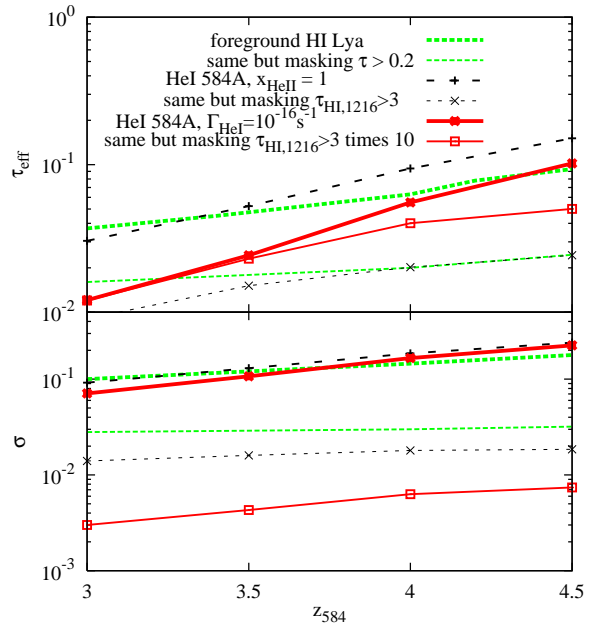, height=8.5cm}}
\end{center}
\caption{Effective optical depth ($\tau_{\rm eff}$) and standard deviation in the transmission ($\sigma$) as a function of the redshift of the $584~$\AA\ absorption ($z_{584}$).  These calculations assume $\Gamma_{\rm HeI} = \Gamma_{\rm HI}$, and $\Gamma_{\rm HI}$ is chosen to match observations of the mean \HI\ Ly$\alpha$ absorption.  The curves represent \HeI\ $584~$\AA\ absorption for the cases  (1) an IGM with ${x}_{\rm HeII} = 1$ (thick long-dashed curve with pluses), (2) the same but masking regions that correspond to $\tau_{\rm HI, 1216} > 3$ in the coeval Ly$\alpha$ forest (thin long-dashed with x's), (3) the \HeI\ $584~$\AA\ line using our model with $\Gamma_{\rm HeII} = 10^{-16}~$s$^{-1}$ in the voids (thick solid with filled squares), and (4) the same but again masking regions that have $\tau_{\rm HI, 1216} > 3$ and \emph{multiplying by 10 in both panels} (thin solid with open squares).  Also shown are $\tau_{\rm eff}$ and $\sigma$ for the foreground Ly$\alpha$ absorption that falls on top of the $584~$\AA\ signal from $z_{584}$ (thick short-dashed curve), and the same but masking regions with $\tau > 0.2$ to suppress this Ly$\alpha$ foreground (thin short-dashed curve).
\label{fig:abs}}
\end{figure}

The top panel in Figure~\ref{fig:abs} plots the effective optical depth (defined as $\tau_{\rm eff} = - \log{\bar{{\cal T}}}$, where $\bar{{\cal T}}$ is the average transmission).  This quantity is plotted for both the \HeI\ $584~$\AA\ forest and the foreground Ly$\alpha$ forest.  The \HeI\ $584~$\AA\ value of $\tau_{\rm eff}$ is shown for the model with ${x}_{\rm HeII} = 1$ (thick dashed black curve) and the model with $\Gamma_{\rm HeII} = 10^{-16}~$s$^{-1}$ (thick solid red curve).  The value of $\tau_{\rm eff}$ is comparable in these two models and comparable to that of the foreground Ly$\alpha$ absorption (thick green dashed curve).  Yet, $\tau_{\rm eff}$ is $3$ times larger at $z=3$ in the ${x}_{\rm HeII} =1$ model compared to the $\Gamma_{\rm HeII} = 10^{-16}~$s$^{-1}$ model, which could potentially allow these models to be distinguished.  Also, the value of $\tau_{\rm eff}$ in the case with $\Gamma_{\rm HeII} = 5\times 10^{-16}~$s$^{-1}$ (not shown in Fig.~\ref{fig:abs}) is a factor of $10$ smaller at $z=3$ than the ${x}_{\rm HeII} =1$ case. If $\Gamma_{\rm HeII}$ fluctuates spatially such with $\Gamma_{\rm HeII} > 5\times 10^{-16}~$s$^{-1}$ (approximately the minimum value derived from \HeII\ Ly$\alpha$ analyses at $z\approx 3$; \citealt{mcquinn09b}), the signal will be even smaller than in this case.
%{\bf \HeII\ forest observations, in combination with \HI\ measurements, derive values for $\Gamma_{\rm HeII}$ (assuming $\Gamma_{\rm HI} = 10^{-12}~$s$^{-1}$) spanning at least the range $5\times 10^{-16}$ --$3\times 10^{-14}~$s$^{-1}$ at $z \sim 2.8$ \citep{zheng04, fechner07}.}

The thin curves that represent  \HeI\ $584~$\AA\ absorption in the top panel in Figure~\ref{fig:abs} are the same as the thick, except regions are masked with $\tau_{\rm HI, 1216} >3$ in the coeval Ly$\alpha$ forest (and the thin curves for the model with $\Gamma_{\rm HeII} = 10^{-16}~$s$^{-1}$ are multiplied by $10$).  This figure also demonstrates that the transmission is much different between the $x_{\rm HeII} = 1$ and $\Gamma_{\rm HeII} = 10^{-16}~$s$^{-1}$ models if denser regions that have $\tau_{\rm HI, 1216} >3$ are masked.

\subsection{Recombining Regions and Double Reionization Scenarios}
We have shown that the amount of \HeI\ absorption is substantially different between the case where the \HeII\ was ionized by a weak background and the case $x_{\rm HeII} = 1$.  There is also the exciting possibility that the \HeII\ was recombining after being ionized by an early generation of sources \citep{venkatesan03} or after a nearby quasar had turned off.  In this case, the \HeII\ fraction of a gas parcel is $x_{\rm HeII} = 1 - \exp(-\Delta t/t_{\rm rec})$, where $\Delta t$ is the time since the \HeII-ionizing background turned off, $t_{\rm rec} = (\alpha_{\rm HeII}^{\rm B} \, n_e)^{-1}$, and $\alpha_{\rm HeII}^{\rm B}$ is the Case B recombination coefficient.  At $z = 3$ and $T = 20,000~$K, $t_{\rm rec}$ is roughly equal to $0.8$ of the Hubble time at the mean density.

Thus, the helium in underdense regions at $z\sim 3$ would have remained doubly ionized for $\Delta t \gtrsim H(z)^{-1}$ after the background turned off.  The density dependence of $x_{\rm HeII}$ is different in this case than in the other cases we have considered, such that there is the possibility that it also can be distinguished using \HeI\ absorption.  However, it takes an extremely weak background with $\Gamma_{\rm HeII} > \alpha_{\rm HeII}^{\rm B} \, n_e \approx  1.2 \times10^{-17}~\Delta_b~$s$^{-1}$ to counteract recombinations in a region. \citet{mcquinn09} found in simulations that such a background developed soon after \HeII\ reionization by quasars was underway.  Therefore, we consider the recombining scenario to be less likely than the others at $3 \lesssim z \lesssim 4$, but nevertheless a tantalizing possibility.
%\footnote{\citet{mcquinn09} found in simulations that the helium in few regions recombined to \HeII\ once \HeII\ reionization was well underway because enough of an almost uniform background of hard, ionizing photons had built up that kept it doubly ionized.} 

\section{Verifying that \HeII\ Reionization was Occurring}
\label{sec:tests}
 
The $z\sim 3$ \HeII\ reionization process is expected to have been very patchy.  If quasars are the source of the ionization, models predict large-scale patches of primarily \HeII\ or \HeIII\ gas that spanned many tens of cMpc \citep{furlanetto08, mcquinn09}.  Even in \HeIII\ regions, the densest gas parcels would have self-shielded such that the helium remains largely singly ionized.  Therefore, a detection of \HeI\ absorption will not necessarily indicate that \HeII\ reionization was occurring, but a detection of \HeI\ absorption from low density gas would.  The lower the density in which this absorption is detected, the stronger the constraint on $\Gamma_{\rm HeII}$ (and $x_{\rm HeII}$) in models where the gas was ionized by a weak background.  Furthermore, the \HI\ Lyman forest provides a measure of the density of an absorber, which allows one to target low density gas parcels.  We showed in Section~\ref{sec:calculations} that the amount of \HeI\ absorption in regions with $\tau_{\rm HI, 1216} <3$ (which roughly correspond to regions with $\Delta_b < 1$) is drastically different between such models.

If absorption from low density gas were detected, it would need to be proven that it is not due to a low-z interloper in order to claim a detection of a large-scale \HeII\ region.  This section discusses two methods to establish that \HeII\ reionization was occurring at $z\sim 3$:  (1) a direct method that involves identifying individual weak \HeI\ lines, and (2) a statistical method that uses the cross correlation with the Ly$\alpha$ forest to detect the \HeI\ absorption.  Unlike the former method, the latter method can be performed even when the signal-to-noise ratio (SNR) on individual \HeI\ lines is much less than unity.

 \subsection{Direct Identification}

With a high enough SNR per resolution element, it may be possible to directly detect individual \HeI\ absorbers as was done in \citet{reimers93} (there, for relatively dense systems).  Once a candidate absorber is identified, it then needs to be ruled out that the absorption is not from a foreground interloper.  To detect \HeII\ gas from regions near the cosmic mean density requires targeting \HeI\ absorbers with $\tau_{\rm HeI, 584} \approx 0.1$ ($\tau_{\rm HI, 1216} \approx 3$).  
Here is an example of how direct identification might be performed:  For wavelengths that correspond to \HeI\ absorption at $z\sim 3$, we count $ \approx 10$ \HI\ Ly$\alpha$ systems in $10^4$~km~s$^{-1}$ of spectrum with a maximum flux decrement between $5$ and $10$ percent.  The probability that one of these systems falls within $50$~km~s$^{-1}$ of a single \HeI\ absorber with the same decrement is $10\%$ and of $2$ \HeI\ absorbers is $1\%$.  Thus, by targeting \HeI\ absorption from coeval \HI\ systems, one can rule out chance contamination from foreground Ly$\alpha$ once a few systems are detected.

An alternative method for direct identification would be to find a second \HeI\ line (such as the $537~$\AA\ line) from the same absorption system.  However, because the absorption in other \HeI\ lines will be at least several times weaker, it will be difficult to detect a second line from systems near the cosmic mean density (requiring SNR~$\sim 100$ per spectral pixel). 

\subsection{Cross Correlation}
\label{ss:cc}
\begin{figure}
\begin{center}
\rotatebox{-90}{\epsfig{file=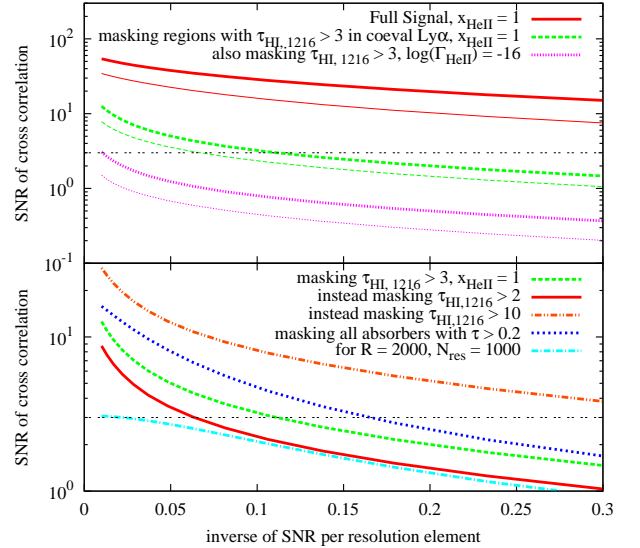, height=10.7cm}}
\end{center}
\caption{Average SNR for detection of \HeI\ absorption in cross correlation.  This quantity is plotted as a function of $\sigma_{\rm inst}$ (the inverse of the average SNR on the continuum in a resolution element), assuming a spectrum with ${\cal R} = 20,000$ and that the measurement uses $1000$ resolution elements ($\Delta z \approx 0.2$).  In the top panel, the thick curves correspond to $z=4$, and the thin to $z=3$.  The solid curves are the full \HeI\ signal, and the lower sets of curves mask overdense regions that have coeval \HI\ Ly$\alpha$ optical depths of $\tau_{\rm HI, 1216} > 3$.  The bottom panel investigates the effect of different masks based on coeval \HI\ absorption or on the observed optical depth $\tau$ in the $584~$\AA\ forest.  Also shown is the SNR of detection at ${\cal R} = 2000$ and for $1000$ resolution elements, again masking regions with coeval $\tau_{\rm HI, 1216} > 3$.\label{fig:matched}}
\end{figure}

It is most likely that the SNR per resolution element will not be high enough to directly identify \HeI\ lines in regions with $\Delta_b \sim 1$ with present instruments.  Since $\tau_{\rm HeI, 584} \propto x_{\rm HeII} \,  \tau_{\rm HI, 1216}$, with $0 < x_{\rm HeII} < 1$ and potentially more structure in  $\tau_{\rm HI, 1216}$ than in $x_{\rm HeII}$, a measurement of $\tau_{\rm HI, 1216}$ may be used to construct a model for the \HeI\ signal that can be used to extract it in cross correlation.  In what follows, we denote the \HeI\ transmission field as $\Trans_{\rm HeI, {\it i}}$, where $i$ labels the spectral pixel.  Furthermore, we assume $\Gamma_{\rm HeI}/\Gamma_{\rm HI}=1$ and $x_{\rm HeII}=1$  to estimate the \HeI\ transmission from $\tau_{\rm HI, 1216}$.  Thus, our estimate for the \HeI\ transmission in a pixel is $\Trans_{\rm HeI, i}^{\rm est} = \exp(-0.025 \,\tau_{{\rm HI, 1216}, {\it i}})$  (cf. equations~\ref{eqn:584GP2}~and~\ref{eqn:tauNHI}).  However, in the limit $\tau_{\rm HeI, 584} \ll 1$, the SNR of the cross correlation is not affected by what we assume for the proportionality between $\tau_{\rm HeI, 584}$ and $\tau_{\rm HI, 1216}$.  In fact, we find that it is only marginally sensitive to this proportionality in realistic cases because the \HeI\ absorption is weak and because the SNR in cross correlation depends more on the phase of modes than on their amplitude.  We discuss in Sec.~\ref{sssec:imperfecttemplate} how an imperfect estimate for $\Trans_{\rm HeI}^{\rm est}$ affects our estimates.  

An estimator for detecting this signal in cross correlation is
\begin{equation}
\widehat{\left( \frac{S}{N} \right)^2} =  \sum_{\forall k} \left( \frac{\tilde{\Trans}_{\rm HeI}^{\rm est}(k)^* \,  \tilde{\Trans}_{\rm HeI}(k)}{|\tilde{\Trans}_{\rm HeI}^{\rm est} (k)|^2} \right) \frac{ \tilde{\Trans}_{\rm HeI}^{\rm est}(k)^* \tilde{D}(k)} {P_{\rm HI}(k) + P_{\rm inst}},  \label{eqn:estimator}\\
\end{equation}
where ${D}$ represents the data, tildes denote a quantity in Fourier space (a convenient basis because the covariance matrix of the noise is diagonal), $P_{\rm HI}$ is the power spectrum of the foreground \HI\ absorption, and $P_{\rm inst}$ is the power spectrum of the instrumental noise (which we assume is white).  Conveniently, the variance on this estimator owing to noise is its average, making it an estimate for the square of the SNR.  This estimator is unbiased in the sense that components that do not correlate with $\tilde{\Trans}_{\rm HeI}^{\rm est}$ do not contribute.  This estimator assumes that the signal is known since it requires $\tilde{\Trans}_{\rm HeI}$ as input.  In practice, this assumption can be avoided  because for interesting cases we find that $\tilde{\Trans}_{\rm HeI}^{\rm est} \propto \tilde{\Trans}_{\rm HeI}$ (and the normalization factor can be determined in a measurement to fractional precision $\approx 0.5$\,SNR$^{-1}$).

The ensemble average of this estimator over the noise and foreground absorption yields
\begin{eqnarray}
\bigg \langle \widehat{\left(\frac{S}{N} \right)^2}  \bigg \rangle 
&=&  \sum_{\forall k} \frac{r(k)^2 \; |\tilde{\Trans}_{\rm HeI}(k)|^2}{P_{\rm HI}(k) + P_{\rm inst}},  \label{eqn:ston}\\
&\sim& N_{\rm res} \, \frac{\sigma_{\rm HeI}^2}{\sigma_{\rm HI}^2 + \sigma_{\rm inst}^2}.\label{eqn:stonapprox}
\end{eqnarray}
The cross correlation coefficient $r(k)$ is bounded such that $-1 <  r(k) < 1$ and is defined as
\begin{equation}
r(k) \equiv \frac{\tilde{\Trans}_{\rm HeI}^{\rm est *}(k) \, \tilde{\Trans}_{\rm HeI}(k)}{( |\tilde{\Trans}_{\rm HeI}^{\rm est}(k)|^2 \, |\tilde{\Trans}_{\rm HeI}(k)|^2)^{1/2}}.
\end{equation}

The approximate equality given by equation (\ref{eqn:stonapprox}) is a rough estimate for the SNR of detection if $r \approx 1$ and if the information is coming from modes near the resolution limit  (which roughly holds for ${\cal R}\lesssim 20,000$).  In this equality, $N_{\rm res}$ is the number of resolution elements, and  $\sigma_{\rm HeI}^2$, $\sigma_{\rm HI}^2$ and $\sigma_{\rm inst}^2$ are respectively the variance of the \HeI\ $584~$\AA\ transmission (normalized such that $\tau_{\rm HeI, 584} =0$ is transmission of unity), the foreground \HI\ transmission, and the instrumental noise in a resolution element.  This approximate relation shows that, for ${\cal R} = 2\times 10^4$ and $N_{\rm res} = 10^3$ ($\Delta z \approx 0.2$), such a cross correlation can be used to detect the signal even if the fluctuations in the signal are more than an order of magnitude smaller than the fluctuations in the noise.\footnote{The discussion in this section could equivalently be phrased in terms of optimal signal extraction with a matched filter, where the matched filter is $M_i \equiv [\bfC_{N}^{-1}]_{ij} \, \Trans_{\rm HeI, j}^{\rm est}$ and $\bfC_N$ is the covariance matrix of the noise (composed of both the instrumental noise and low-z Ly$\alpha$ absorption).  For $\Trans_{\rm HeI, j}^{\rm est} = \Trans_{\rm HeI, j}$, $M_i$ is the filter that, when its dot product is taken with the signal, maximizes in the filtered data the ratio of power in the signal to the average power in the noise.  Furthermore, equation (\ref{eqn:ston}) is the SNR that such a filter can achieve.}

Figure~\ref{fig:matched} plots the expected average SNR as a function of $\sigma_{\rm inst}$ at which the \HeI\ absorption can be detected in cross correlation, assuming $N_{\rm res} = 10^3$ and ${\cal R} = 2\times 10^4$.  To calculate this quantity, we substitute terms of the form $\tilde{\Trans}^\dag \tilde{\Trans}$ with their ensemble-average in equation (\ref{eqn:ston}).    The red curves are this SNR for observations of \HeI\ absorption at $z=3$ (thin) and $z=4$ (thick).   \HeI\ absorption can be detected statistically using this technique even with an extremely poor SNR per resolution element (at $\approx 10\,\sigma$ for $\sigma_{\rm inst} = 0.3$).  In the case of COS, $N_{\rm res} \approx 500, 1000$, and $1500$, depending on whether $1,2$, or $3$ of the $\approx 40$~A bandpasses (that each COS NUV grating can simultaneously observe) cover the \HeI\ forest for a given target.

\begin{figure}
\begin{center}
\rotatebox{-90}{\epsfig{file=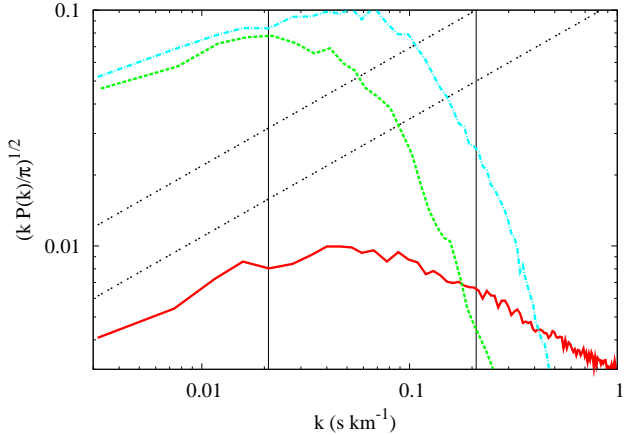, height=8.5cm}}
\end{center}
\caption{Square root of the dimensionless absorption power spectrum.  The dot-dashed cyan curve is this for the \HeI\ absorption at $z=4$, and the thick dashed green is this for the foreground Ly$\alpha$ absorption.  The solid red curve is this for the $z=4$ \HeI\ signal after masking regions with coeval $\tau_{\rm HI, 1216} > 3$ (the different scaling at high-$k$ compared to the blue curve owes to the mask).  The vertical lines are the Nyquist cutoff for ${\cal R} = 2000$ and ${\cal R} = 20,000$.  
The black dotted diagonal lines are the instrumental noise assuming ${\cal R} = 20,000$ and either $\sigma_{\rm inst} = 0.05$ or $\sigma_{\rm inst} = 0.1$.  
\label{fig:pk}}
\end{figure}

For the calculations in Figure~\ref{fig:matched}, wavevectors only up to the Nyquist cutoff for the spectrograph resolution are included  ($k_{\rm N} \equiv \pi {\cal R}/c$) , with this cutoff imposed using a tophat filter in $k$-space.  Spectrographs typically sample the signal $S(n) = \int d\lambda' \, W_{\cal R}(n \, \Delta \lambda - \lambda')\, {\cal F}(\lambda')$ to wavevectors that are a few times higher than $k_{\rm N} $ to avoid aliasing (i.e., $\lambda/\Delta \lambda \sim {\rm few} \, {\cal R}$), where $n$ is the integer number of the spectral pixel, ${\cal F}$ is the incident flux, and $W_{\cal R}$ is the detector window function that damps wavevectors above $k_{\rm N}$.  For realistic $\sigma_{\rm inst}$ and ${\cal R} \gtrsim 20,000$, most contributions to the \HeI\ SNR of detection are coming from $k$-modes a couple times smaller than the Nyquist wavevector (from wavevectors where $P_{\rm inst} \approx P_{\rm HI}$), such that our results are somewhat robust to the exact details of the spectrograph.  This can be noted in Figure~\ref{fig:pk}, which features the power spectrum of the different components of the signal.  The rightmost vertical line is the Nyquist cutoff for ${\cal R} = 20,000$.  The foreground Ly$\alpha$ power spectrum (dashed green curve) intersects the noise power for $\sigma_{\rm inst} \sim 0.05$ and $\sigma_{\rm inst} \sim 0.1$ (dotted black diagonal lines) prior to the Nyquist cutoff.

This point of intersection also shows that both the instrumental noise and the noise from foreground absorption are of comparable importance in determining the SNR.  Interestingly, because of the different thermal broadening scales, the \HeI\ absorption has significantly more power than the foreground absorption at $k \gtrsim 0.1~$s~km$^{-1}$ (cyan dot-dashed curve, Fig. \ref{fig:pk}).  This difference increases the detectability of the \HeI\ absorption in ${\cal R} \gtrsim 20,000$ spectra.  The \HI\ foreground absorption is also highly non-Gaussian, which could affect the probability distribution of the estimated SNR from equation (\ref{eqn:estimator}), especially for short skewers that do not sample a representative set of foreground absorption lines.  We have calculated the probability distribution of the estimated SNR from mock observations constructed from our simulations.  In practice we find that the probability distribution is relatively Gaussian even for $25/h$~cMpc skewers.

\subsubsection{Masking to measure \HeI\ absorption in low density regions}

To isolate the \HeI\ absorption from low density gas, we mask pixels in the simulated \HeI\ forest with coeval Ly$\alpha$ absorption of $\tau_{\rm HI, 1216} > 3$.  This mask covers $10\%$ of the pixels at $z=3$ and $25\%$ at $z=4$.  The calculation of the SNR in the presence of a mask is more complicated and is no longer diagonal in Fourier space.  In addition, the mask artificially correlates noise wavevectors with those of the signal.  To avoid such issues, our calculations only mask the signal (by setting $\tau_{\rm HI, 1216} = 0$) and not the noise to calculate the SNR via equation (\ref{eqn:estimator}).  This estimate for the SNR likely underestimates the value that an optimal analysis could achieve because it involves adding additional noise to the signal.  (See Appendix \ref{ap:SNRcc} for additional discussion with regard to the effect of a mask.)

The full calculation in the presence of masks is computationally tractable (with just the additional complication of inverting $N_{\rm pixel} \times N_{\rm pixel}$ matrixes, where $N_{\rm pixel}$ is the number of spectral pixels).  In addition, a simplified analysis that applies the cross-correlation statistic (eqn.~\ref{eqn:estimator}) to just the intervals between masked regions, and sums the SNR$^2$ from each of these intervals, would likely provide a reasonable estimate for the SNR of detection.

For the mask with $\tau_{\rm HI, 1216} > 3$, the standard deviation of the \HeI\ absorption is reduced by approximately a factor of $10$ for the model with ${x}_{\rm HeII} = 1$ relative to the unmasked model, and by a factor of more than $100$ for the model with $\Gamma_{\rm HeII} = 10^{-16}~$s$^{-1}$ (Fig. \ref{fig:abs}).  However, even when masking $\tau_{\rm HI, 1216} > 3$, it is still possible to detect the signal in the model with ${x}_{\rm HeII} = 1$  with $N_{\rm res} \sim 10^3$ for $\sigma_{\rm inst} = 0.1$ (the dashed green curve in the bottom panel of Fig. \ref{fig:matched}).  This value $\sigma_{\rm inst} = 0.1$ may be achievable with COS on several existing sightlines, as discussed in the introduction.

Thus far we have discussed masking regions that correspond to $\tau_{\rm HI, 1216} > 3$.  Such a mask requires a SNR~$\approx 50$ on the \HI\ Ly$\alpha$ forest continuum at ${\cal R} \geq 2\times 10^4$ to be able to detect flux in pixels with $\tau_{\rm HI, 1216} = 3$ at $2 \, \sigma$.  Such a SNR may not be achievable on all \HeI\ forest quasars.  If instead only SNR~$\approx 20$ can be achieved, then masking $\tau_{\rm HI, 1216} > 2$ is more reasonable.  The signal can still be detected at $3 \, \sigma$ with such a mask with $\sigma_{\rm inst} \approx 0.06$,  ${\cal R} = 20,000$, and $N_{\rm res} = 1000$, as illustrated by the red solid curve in the bottom panel of Figure~\ref{fig:matched}.  In addition, if Ly$\beta$ information is used, masking only regions with $\tau_{\rm HI, 1216} > 10$ is even possible.  Such a mask results in much more signal for the ${x}_{\rm HeII} = 1$ model (orange dot-dashed curves), with a $3 \, \sigma$ detection possible even at $\sigma_{\rm inst} \approx 0.3$.  Even though this mask includes denser regions than the mask with $\tau_{\rm HI, 1216} > 3$, the expected SNR is a factor of $3$ times larger at $\sigma_{\rm inst} \approx 0.1$ in the ${x}_{\rm HeII} = 1$ model than in the model with $\Gamma_{\rm HeII} = 10^{-16}~$s$^{-1}$.\footnote{The data can be coaxed further to improve the significance of detection.  For example, much of the contamination from foreground \HI\ absorption can be eliminated by removing all pixels that have $\tau > 0.2$ (see the thin green dashed curves in Fig. \ref{fig:abs}).  The motivation for this additional mask is that the foreground absorption is characterized by rare absorbers that typically have larger optical depths than the \HeI\ absorbers.   This operation improves the SNR further (blue dotted curve in the bottom panel of Fig. \ref{fig:matched}).  Although, such an operation would make the full analysis more complicated because it correlates the foregrounds with the mask.   In addition, masking lines based on their associated Lyman-series and other metal lines would further improve the SNR.}

\subsubsection{Imperfect signal template estimates}
\label{sssec:imperfecttemplate}

The exact form of the \HeI\ absorption signal will never be known, and instead we have some best guess $\Trans_{\rm HeI, i}^{\rm est}$ for its form from coeval Ly$\alpha$ forest measurements.  In cases where the signal is not perfectly known, the achievable SNR is decreased by the cross correlation coefficient $r$ (eqn.~\ref{eqn:ston}).  The different amounts of thermal broadening between \HeI\ and \HI\ absorption lines make $r < 1$ even with a perfect measurement of $\tau_{\rm HI, 1216}$.  This effect was included in all our calculations, but we find it has a negligible effect on the SNR for detection of the cross correlation ($\lesssim 10\%$).  To the extent the IGM had a single temperature, the effect of thermal broadening is a convolution in wavelength space with a single Gaussian filter.  In this single-temperature limit, the filters divide out and do not affect the value of $r$ and, therefore, the SNR of detection.  

Another complication arises because the measurement of $\tau_{\rm HI, 1216}$ will never be perfect.  There will be noise in the estimate for this field that reduces $r$ further.  To test the importance of an imperfect reconstruction of $\tau_{\rm HI, 1216}$, we have added noise to $\tau_{\rm HI, 1216}$ with the standard deviation in the noise per pixel of $\sigma^{\rm HI}_{\rm inst}/\exp(-\tau_{\rm HI, 1216})$.  We find a negligible difference in the SNR of detection for $\sigma^{\rm HI}_{\rm inst} \lesssim 0.1$ for the case with the $\tau_{\rm HI, 1216} > 3$ mask, which indicates that noise in the reconstruction of $\tau_{\rm HI, 1216}$ does not significantly degrade the measurement.\footnote{We found earlier that noise at the $10\%$-level in the NUV spectrum can have a substantial effect on the SNR at which the \HeI\ absorption is detected.  The reason that a similar level of noise is less important in the \HI\ spectrum is because $\tau_{\rm HI, 1216} \sim 1$ for most pixels at these redshifts, which translates into a $\sim 10\%$ error on $\tau_{\rm HI, 1216}$ if $\sigma^{\rm HI}_{\rm inst} = 0.1$, except in the low optical depth pixels (where there is little \HeI\ signal anyway) and the high ones.  Whereas for \HeI\ absorption, if $\sigma^{\rm HeI}_{\rm inst} = 0.1$, we have order unity uncertainty in this optical depth for typical $\tau_{\rm HeI, 584}$.  Furthermore, $r$ (and thus the SNR of detection) is most sensitive to the phase of modes, which is largely maintained even when significant noise is added to the high-$\tau_{\rm HI, 1216}$ pixels.}

The \HeI\ absorption signal can also differ from the estimate for this signal from $\tau_{\rm HI, 1216}$ because of the patchy structure of \HeII\ reionization.  The signal in the cross correlation is coming from $k \sim 1~$cMpc$^{-1}$, whereas the structure in the \HeII\ fraction is thought to be modulated at scales of $R_{\rm bubble} \sim 30$~cMpc, the size of the \HeIII\ bubbles around quasars \citep{furlanetto09, mcquinn09}.  The wavevectors affected by the structure of \HeII\ reionization $k \sim R_{\rm bubble}^{-1}$ will be more than an order of magnitude smaller than those that contribute to the SNR.  If HeII reionization is perfectly patchy such that the helium is either singly ionized or doubly ionized and a skewer of length $\gg R_{\rm bubble}$ is measured, then $r \approx \bar{x}_{\rm HeII}^{1/2}$ at $k \gg  R_{\rm bubble}^{-1}$, where $\bar{x}_{\rm HeII}$ is the volume-averaged \HeII\ fraction.   Therefore, the total SNR of the cross correlation is reduced on average by the factor $\bar{x}_{\rm HeII}$.  Also note that in the contrasting toy case that \HeII\ reionization was perfectly homogeneous, the SNR is also reduced by $\bar{x}_{\rm HeII}$.

A detection of \HeI\ absorption from low density gas would translate to a $1 \, \sigma$ error on $\bar{x}_{\rm HeII}$ of $\delta x_{\rm HeII} \approx \bar{x}^e_{\rm HeII}({\rm {SNR}})/{\rm {SNR}}$, where $\bar{x}^e_{\rm HeII}({\rm SNR})$ is the expected $\bar{x}_{\rm HeII}$ for a given estimated SNR of detection in cross correlation with $\exp(-0.025\,\tau_{\rm HI, 1216})$.  Thus, a $4\,\sigma$ detection from low density gas would yield a $25\%$ constraint (ignoring uncertainty in $\Gamma_{\rm HeI}$).  This constraint assumes that the low density gas during \HeII\ reionization was composed of large-scale regions in which the helium was either \HeII\ or \HeIII, which is expected theoretically and is what is seen in simulations \citep{mcquinn09}. 

\section{Conclusions}
This paper discussed the usefulness of the \HeI\ $584~$\AA\ forest to study \HeII\ reionization at $3 \lesssim z \lesssim 4.5$.  The optical depth of the \HeI\ $584~$\AA\ line is proportional to the optical depth in \HI\ Ly$\alpha$ by the factor $0.025 \, x_{\rm HeII} \, \Gamma_{\rm HI}/\Gamma_{\rm HeI}$ (ignoring differences in the amount of thermal broadening between \HI\ and \HeI).  The factor $x_{\rm HeII}$ should have been spatially variable, but the ratio $\Gamma_{\rm HI}/\Gamma_{\rm HeI}$ should have been essentially spatially independent and roughly equal to unity at relevant redshifts.   Therefore, this absorption can be used to study \HeII\ reionization through its dependence on $x_{\rm HeII}$.  

Our best method at present to probe \HeII\ reionization, \HeII\ Ly$\alpha$ absorption, saturates at neutral fractions of a part in a thousand at the cosmic mean density.  In contrast, \HeI\ $584~$\AA\ absorption is unsaturated in all except the densest regions, even for $x_{\rm HeII} = 1$.  This absorption provides a complementary window into the ionization state of intergalactic helium at $z\sim 3$ that can definitively test whether an opaque region in the \HeII\ Ly$\alpha$ forest was due to a large-scale \HeII\ region.  Even for realistic amounts of instrumental noise and foreground \HI\ Ly$\alpha$ absorption, we showed that the coeval \HI\ Ly$\alpha$ forest absorption can be used to construct a matched filter that can detect the \HeI\ absorption at high significance from a single quasar sightline.

A detection of \HeI\ absorption from $\Delta_b \sim 1$ gas at $z \sim 3$ would definitively indicate that \HeII\ reionization was occurring.   If a significant fraction of the intergalactic helium was in \HeII, we found that \HeI\ $584~$\AA\ forest absorption from low density gas could be identified in a quasar spectrum with SNR~$\sim10$ and ${\cal R} \sim 10^4$ in an interval of $\Delta z = 0.2$.  These specifications may be achievable with the COS instrument on the HST for a few known $z>3$ targets. \\ 

We thank Claude-Andr{\'e} Faucher-Gigu{\`e}re, Wayne Hu, Gabor Worseck, and especially J. Xavier Prochaska for useful discussions.   We also thank Claude-Andr{\'e} Faucher-Gigu{\`e}re, Lars Hernquist, Adam Lidz, and Volker Springel for providing the simulations used in this work.  E. S. acknowledges support by NSF Physics Frontier Center grant PHY-0114422 to the
Kavli Institute of Cosmological Physics. \\

\bibliographystyle{apj}

\begin{appendix}

\section{A. Measuring the Hardness of the Ionizing Background with \HeI\ Absorption}
\label{sec:hardness}

We argued in Section \ref{sec:forestGamma} that the spectrum between $13.6~$eV and \HeII\ Ly$\alpha$ at $40.8~$eV (and, thus, the ratio ${\Gamma_{\rm HI}}/{ \Gamma_{\rm HeI}}$) depends only on the \HI\ column density distribution and the spectrum of the sources.   Since the \HI\ column density distribution is constrained by observations, a measurement of ${\Gamma_{\rm HI}}/{ \Gamma_{\rm HeI}}$ will constrain the spectrum of the sources in this band \citep{miralda92, santos03}.  This ratio can be measured by comparing the continuum absorption at $912 \, (1+z)~$\AA\ -- the \HI\ limit -- and $504\,(1+z)~$\AA\ -- the \HeI\ limit \citep{miralda92}.  These two optical depths are related via
\begin{eqnarray}
\tau_{\rm HeI}^{\rm cont}(E^{\rm ion}_{\rm HI}) &\approx& \,x_{\rm HeII} \, \frac{n_{\rm He}}{n_{\rm HI}} \, \frac{\alpha_{\rm HeI}}{\alpha_{\rm HI}} \frac{\Gamma_{\rm HI}'}{ \Gamma_{\rm HeI}'} \, \frac{\sigma_{\rm HeI} (E^{\rm ion}_{\rm HeI})}{ \sigma_{\rm HI} (E^{\rm ion}_{\rm HI})}  \,  \tau_{\rm HI}^{\rm cont}(E_{\rm ion}^{\rm HeI}), \nonumber \\
& \approx & 0.10 \,x_{\rm HeII}  \,\frac{\Gamma_{\rm HI}'}{ \Gamma_{\rm HeI}'}\, \tau_{\rm HI}^{\rm cont}(E^{\rm ion}_{\rm  HeI}),
\end{eqnarray}
where primes denote the average value for the quantity inside the absorber.  This ratio of continuum optical depths can be measured most easily in systems with $N_{\rm HI} \sim 10^{17}~$cm$^{-2}$, corresponding to systems with  $\tau_{\rm HI}^{\rm cont}(E^{\rm ion}_{\rm  HeI}) \sim 1$.    These column densities will be extremely self-shielded to \HeII-ionizing photons such that $x_{\rm HeII} = 1$ is a good assumption.  Thus, the ratio of the continuum optical depths can be used to measure $\Gamma_{\rm HI}'/ \Gamma_{\rm HeI}'$.  Because this method relies on detecting smooth spectral features, this science can be pursued even in much lower resolution spectra than are required to study \HeII\ reionization, and a $< 10\%$ constraint on the continuum is required to detect the weak breaks from \HeI\ continuum absorption in systems with $\tau_{\rm HI}^{\rm cont}(E^{\rm ion}_{\rm  HeI}) \sim 1$ and, thereby, constrain $\Gamma_{\rm HI}'/ \Gamma_{\rm HeI}'$.

The primary complication with this continuum-absorption method to constrain ${\Gamma_{\rm HI}}/{ \Gamma_{\rm HeI}}$ is that \HeI\ self-shields at \HI\ column densities that are a decade larger than those for which the \HI\ self-shields.  Thus,  in systems with $\tau_{\rm HI}^{\rm cont}(E^{\rm ion}_{\rm  HeI}) \sim 1$, $\Gamma_{\rm HI}' < \Gamma_{\rm HI}$ even though $\Gamma_{\rm HeI}' \approx \Gamma_{\rm HeI}$.   However, if the absorber is a slab of material for which both the line of sight and the incident radiation is perpendicular to the plane of the slab, then $\Gamma_{\rm HI}' = \Gamma_{\rm HI} \, [1- \exp({-\tau^{\rm cont}_{\rm HI}})]/\tau^{\rm cont}_{\rm HI}$.  The difference in this case between $\Gamma_{\rm HI}'$ and $\Gamma_{\rm HI}$ is relatively small for $\tau^{\rm cont}_{\rm HI} \lesssim 1$.  (In fact, under most circumstances, the correction will be smaller than in this toy example because there will be other orientations with smaller optical depths to locations within the absorber.)  Therefore, a measurement of $\tau_{\rm HeI}^{\rm cont}/\tau_{\rm HI}^{\rm cont}$  (which yields $\Gamma_{\rm HI}'/ \Gamma_{\rm HeI}'$) for $N_{\rm HI} \sim 10^{17}~$cm$^{-2}$ can be mapped to a constraint on $\Gamma_{\rm HI}/ \Gamma_{\rm HeI}$ without significant error.
%\footnote{Another complication is that for $\tau^{\rm cont}_{\rm HI} \gtrsim 1$, reabsorption of photons produced in recombinations directly to the ground state occurs within the system.  This can alter the above relation (which assumes Case A recombination) by a maximum of $40\%$.}

A second method to constrain ${\Gamma_{\rm HI}}/{ \Gamma_{\rm HeI}}$ is to measure the equivalent width of a line in the \HeI\ forest and to compare this with the equivalent width of this system's \HI\ Lyman-series absorption lines \citep{santos03}.  This method can be employed on systems that have much smaller \HI\ column densities than the above continuum absorption method.  What this method requires is absorption systems that have $x_{\rm HeII} \approx 1$ in order to directly measure $\Gamma_{\rm HI}/ \Gamma_{\rm HeI}$ (equation~\ref{eqn:tauNHI}). 
For $\Gamma_{\rm HeII} = 10^{-15}~$s$^{-1}$, systems with $N_{\rm HI} \gtrsim 10^{15}~$cm$^{-2}$ at relevant $z$ self-shield to \HeII-ionizing photons and can have $x_{\rm HeII} \approx 1$  \citep{mcquinn09}.  At lower $\Gamma_{\rm HeII}$, even smaller \HI\ column densities self shield. (Note that, for a fixed $\Gamma_{\rm HeII}$, the \HeII\ is likely to transition from ionized to neutral over a relatively narrow range in $N_{\rm HI}$.  The same does not happen to the hydrogen once it starts to self-shield, and it is like to remain highly ionized for $N_{\rm HI}\lesssim 10^{19}~$cm$^{-2}$ at $z\sim 3$.)

The major difficulty with the second method is that it will be challenging to measure the \HI\ column density for systems that have a significant optical depth in \HeI\ $584$~\AA.  The use of higher \HI\ Lyman series resonances, or continuum absorption for the largest column densities, will be required for such measurements.  For example, systems with $N_{\rm HI} \sim 10^{15}~$cm$^{-2}$ and $x_{\rm HeII} = 1$ have $\tau_{\rm HeI, 584} \approx 1$ and \HI\ Ly$\gamma$ absorption of $\tau_{\rm HI, 973} \approx 2$, assuming a linewidth of $30$~km~s$^{-1}$.  Even if $x_{\rm HeII} = 1$ is not assumed, such a measurement would place an upper bound on ${\Gamma_{\rm HI}}/{ \Gamma_{\rm HeI}}$.

\section{B. Signal-to-noise of Cross Correlation}
\label{ap:SNRcc}

Denote the measurement and estimated templates with subscript $m$ and $e$, respectively.  Both the template and measurement have some underlying signal and noise, which we represent here for the measurement as $\Trans_m(k)=\Trans_{m,t}(k)+n_m(k)$, where $\Trans_{m,t}$ is the underlying \HeI\ signal and $n_m(k)$ is the measurement noise.  A similar decomposition applies for the estimated signal: $\Trans_e(k)=\Trans_{e,t}(k)+n_e(k)$, where $\Trans_{e,t}(k)$ is the component of the estimate that correlates with the signal and $n_e(k)$ is the component that does not.
The covariance of the cross-power between the \HeI\ measurement and the template estimate, $H_{m,e}(k)$, is then
\begin{eqnarray}
\langle \delta H_{m,e}(k_1) \delta H_{m,e}^*(k_2)\rangle &=& \langle H_{m,e}(k_1) H_{m,e}^*(k_2)\rangle-\langle H_{m,e}(k_1)\rangle\langle H^*_{m,e}(k_2)\rangle, \nonumber \\
&=& \langle \Trans_m(k_1) \Trans_e^*(k_1) \Trans_m^*(k_2) \Trans_e(k_2) \rangle - \langle \Trans_m(k_1) \Trans_e^*(k_1) \rangle \langle \Trans_m^*(k_2) \Trans_e(k_2) \rangle, \nonumber \\
&=& \langle [\Trans_{m,t}(k_1)+n_m(k_1)] [\Trans_{e,t}^*(k_1)+n_e^*(k_1)] [\Trans_{m,t}^*(k_2)+n_m^*(k_2)] [\Trans_{e,t}(k_2)+n_e(k_2)] \rangle, \nonumber \\ & & - \langle [\Trans_{m,t}(k_1)+n_m(k_1)] [\Trans_{e,t}^*(k_1)+n_e^*(k_1)] \rangle \langle [\Trans_{m,t}^*(k_2)+n_m^*(k_2)] [\Trans_{e,t}(k_2)+n_e(k_2)] \rangle, \nonumber \\
&=& \Trans_{m,t}(k_1) \Trans_{m,t}^*(k_2) \langle n_e^*(k_1) n_e(k_2) \rangle + \Trans_{e,t}^*(k_1) \Trans_{e,t}(k_2) \langle n_m(k_1) n_m^*(k_2) \rangle + \langle n_m(k_1) n_m^*(k_2) \rangle \langle n_e^*(k_1) n_e(k_2) \rangle, \nonumber \\
&=& \delta_{k_1,k_2} [ |\Trans_{m,t}(k_1)|^2 N_e(k_1) + |\Trans_{e, t}(k_1)|^2 N_m(k_1) + N_m(k_1) N_e(k_1) ],
\label{eqn:appcrosscovariance}
\end{eqnarray}
where $N_X(k) =  \langle n_X(k) n_X^*(k) \rangle$ and in general, $\Trans_{X,t}(k_1) \Trans_{X,t}^*(k_2)$ is non-diagonal, but is multiplied here by $\langle n_X^*(k_1) n_X(k_2) \rangle$, which is $\delta_{k_1,k_2} N_X(k_1)$.  The significance with which we can measure the amplitude of the cross-power is
\begin{eqnarray}
SNR^2 
&=& \sum_{k_1,k_2} \Trans_{e,t}(k_1) \Trans_{m,t}^*(k_1) \langle \delta H_{m,e}(k_1) \delta H_{m,e}(k_2)\rangle^{-1} \Trans_{e,t}^*(k_2) \Trans_{m,t}(k_2), \nonumber \\
&=& \sum_{k} \frac{|\Trans_{e,t}(k) \Trans_{m,t}^*(k)|^2}{[ |\Trans_{m,t}(k)|^2 N_e(k) + |\Trans_{e,t}(k_1)|^2 N_m(k) + N_m(k) N_e(k) ]}, \nonumber \\
&\approx& \sum_{k} \frac{|\Trans_{e,t}(k) \Trans_{m,t}^*(k)|^2}{|\Trans_{e,t}(k_1)|^2 N_m(k) },
\label{eqn:apsnr}
\end{eqnarray}
where the last line assumed that template estimate noise $N_e(k)$ is negligible.  Ignoring the additional \HeI\ ionization structure from fluctuations in $x_{\rm HeII}$ (which we argued in Section \ref{ss:cc} mostly affects the amplitude of small-scale modes in $\Trans_{m,t}$), one has $N_m(k) = P_{\rm inst}(k)+P_{\rm HI}(k)$.  With this replacement, Eq.~\ref{eqn:apsnr} coincides with Eq.~\ref{eqn:ston}.

In the text, we estimated the SNR in the case where only the signal is masked but not the noise.  The derivation of Eq.~\ref{eqn:apsnr} holds in this case, except that $\Trans_{e,t}(k) \Trans_{m,t}^*(k)$ is replaced with this cross correlation between the (noiseless) estimated and measured signal, including the masked regions.  Furthermore, one could imagine adding mock noise in masked regions to real observations to replicate this case.  

In a proper analysis of observations, the noise will be masked along with the signal.  Since the mask covers a relatively small fraction of space at relevant redshifts, the estimate for the SNR given by Eq.~\ref{eqn:apsnr} should approximately hold.  This can be noted by first considering the SNR one can achieve if the segments between masks were considered separately.  In this case, the SNR$^2$ is the sum of Eq.~\ref{eqn:apsnr} computed from each segment.  The difference in this case is (1) that the SNR is decreased by $\approx (1-f_{\rm mask})^{1/2}$ where $f_{\rm mask}$ is the mask covering fraction because there are $f_{\rm mask}$ fewer $k$ values, and (2) $\Trans_{e,t}(k) \Trans_{m,t}^*(k)$ in Eq.~\ref{eqn:apsnr} is replaced with its value in the unmasked regions (which is slightly different than the replacement discussed in the previous paragraph, which included the power from the mask).  This argument ignores the correlations that occur between the noise from the foreground \HI\ absorption in different segments.  However, these correlations should be small since the low-z forest is characterized by discrete, largely uncorrelated lines.

\end{appendix}

\end{document}